\documentclass[runningheads]{llncs}

\usepackage[utf8]{inputenc}
\usepackage[english]{babel}
\usepackage{graphicx}
\usepackage{amssymb}
\usepackage{amsmath}
\usepackage{float}
\usepackage{xcolor}
\usepackage{hyperref}
\usepackage[T1]{fontenc}
\usepackage{adjustbox}
\usepackage{colortbl}
\usepackage{booktabs}
\usepackage{bbding}

\title{On the similarities and differences between the Cloud, Fog and the Edge
\thanks{This work is supported by the European Union through the Horizon 2020 research and innovation programme under grant 730929.}
}
\author{Sašo Stanovnik\inst{1}\Envelope \and Matija Cankar \inst{1}}
\authorrunning{Stanovnik and Cankar}
\institute{XLAB Research\\XLAB d.o.o.\\Ljubljana, Slovenia\\
\email{\{saso.stanovnik,matija.cankar\}@xlab.si}}

\begin{document}
    \maketitle

    \begin{abstract}
        The field of edge and fog computing is growing, but there are still many inconsistent and loosely--defined
        terms in current literature.
        With many articles comparing theoretical architectures and evaluating implementations, there is a need to
        understand the underlying meaning of information condensed into fog, edge, and similar terms.
        Through our review of current literature, we discuss these differences and extract key characteristics for
        basic concepts that appear throughout.
        The similarities to existing IaaS, PaaS and SaaS models are presented, contrasted against similar models
        modified for the specifics of edge devices and workloads.

        We also evaluate the different aspects existing evaluation and comparison works investigate, including the
        compute, networking, storage, security, and ease--of--use capabilities of the target implementations.
        Following that, we make a broad overview of currently available commercial and open--source platforms
        implementing the edge or fog paradigms, identifying key players, successful niche actors and general trends for
        feature--level and technical development of these platforms.
    \end{abstract}

    \section{Introduction}\label{sec:introduction}
    Computing resources can be made available in a number of ways.
    Using the \emph{grid} as a somewhat low--level abstraction predates using the \emph{cloud} as a more high--level
    abstraction to computing resources, and the latter is, at the moment, the most popular method of delegating
    computational resources.
    With consumer--focused and low--powered machines becoming capable of an increasing number of non--trivial tasks,
    there is often a desire to take advantage of that capacity to perform computation more optimally, e.g.\ by
    increasing data locality.

    Extending from the cloud, edge computing and, more recently, fog computing have appeared as terms for describing
    such architectures.
    The differences are not immediately apparent however, and different sources have sometimes subtly, sometimes very
    prominently, differing view on what each should encompass.

    Our goals are to identify similarities and differences in the approaches to handling the different layers of
    the fog/edge computing architectures and to make a comparison of existing research and solutions.
    We will begin by exploring and clarifying commonly used terms in Section~\ref{sec:commonly-used-terms}, then
    continue on to explore the levels of overall management of platforms in literature in
    Section~\ref{sec:levels-of-management-of-edge-architectures}.
    Following that, we provide an overview of current literature in Section~\ref{sec:existing-research} and conclude
    with an overview of currently available proprietary and open--source platforms in
    Section~\ref{sec:existing-platforms}.

    \section{Commonly used terms}\label{sec:commonly-used-terms}
    Many terms used in this field are used frequently, without a clear consensus on what, specifically, they mean.
    This complicates interpreting existing literature, where different researchers have slightly different views on
    the boundaries between layers, which, while always overlapping, can differ significantly.

    Most of these terms are not technical and are used primarily in marketing and, even there, are established to
    varying degrees.
    They condense many technical details into a well--recognised word that is a generally correct description of the
    concept, but lacks the specificity needed to recognise and understand the issue in--depth.

    \subsection{The Cloud}\label{subsec:the-cloud}
    Although a term that is already established, it is useful to define the key characteristics that researchers in the
    field describe with this term, as not all may be immediately obvious when compressed into a single word.

    Mainly the term concerns the abstraction of physical or virtual resources, made available through a managed
    interface.
    The location, specific configuration and ownership of the resources themselves should not matter, other than for
    their performance characteristics or for differences in billing models.

    \subsection{Fog, Edge and IoT}\label{subsec:fog-edge-and-iot}
    These are terms widely used, but problematic in terms of overlap.
    The somewhat recently emerged field of cloud IoT providers sometimes covers all three aspects, seldom
    well--defined and mostly used interchangeably.
    The research into these platforms exhibits the same, with authors' interpretation implicit and specific to a single
    application.

    This work attempts to non--authoritatively define the scope of each of these terms, based on our cumulative
    understanding of the field rather than based on any specific research.
    The basic constraint is that we attempt to classify devices into a single category, which provides multiple
    opportunities to evaluate differences and overlaps.

    In a larger picture, the \emph{fog}, \emph{edge} and \emph{IoT} layers, joined by the \emph{cloud}, form a
    hierarchical relationship with a single, likely distributed cloud at the top, followed by multiple fogs, each
    containing multiple edge devices, connected to even more IoT devices.

    The only layer the standalone existence of which does not make sense is the fog layer, as this is, usually, a
    bridging layer connecting multiple edge devices and the cloud and, almost by definition, must include edge devices.
    IoT devices are of little use on their own and edge computing--capable devices, on their own, already exist in the
    form of personal computers.

    \subsection{Proposed understanding of the architecture}\label{subsec:proposed-understanding-of-the-architecture}
\begin{figure}
        \centering
        \includegraphics[width=0.95\linewidth]{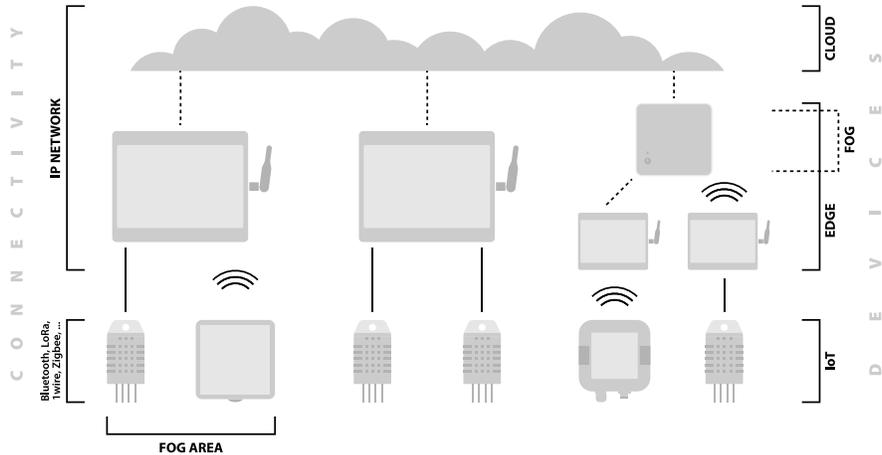}
        \caption{Hierarchical architecture summary.}
        \label{fig:cloud-fog-edge-iot-diagram}
    \end{figure}

    To try to establish an understanding and to define terms used in subsequent sections, we present our definition
    of the above terms.
    This layout is summarised in Figure~\ref{fig:cloud-fog-edge-iot-diagram}.

    Starting from the bottom layer of IoT devices, these are commonly defined as sensors and actuators that interact
    with the environment.
    The definition we use is slightly stricter in that we require them to not run a traditional operating system.
    This excludes powerful single board computers such as the Raspberry Pi, but includes microcontrollers, simple
    sensors communicating over one wire and Bluetooth--based sensor packages.

    This is the first instance of overlap between the IoT and edge layers.
    The Raspberry Pi in particular, having GPIO pins available to connect to the physical world is often considered an
    IoT device because of that fact, but can also act as an edge device.
    The differentiating factor is only the software that runs on them and, subsequently, the role they take in the
    overall architecture.

    Edge devices are, in our definition, devices capable of IP--based networking, running an operating system offering
    remote configuration, connectivity, and being able to run applications on--demand.
    They also connect to devices hierarchically below them, possibly using non--IP--based networks such as
    Bluetooth\cite{bluetoothsiginc.bluetooth} or Thread\cite{threadgroupthread}.
    Apart from bridging different connections, providing computational power is a major role of these devices because
    of them being relatively computationally powerful.
    Examples of such devices are single board computers, laptops and industrial gateways.
    Mobile phones could also be considered edge devices, but as they are not configurable enough to be equivalent to
    others, we do not include them in this definition.

    The most ambiguously defined is the layer of fog devices.
    It is very similar to the edge layer and can be viewed as its vertical extension---there is very little difference
    between the edge and the fog.
    As it is used in current literature, it most often is a general--purpose term used for devices between the topmost
    cloud layer and another layer below them, with the primary differentiator being their primary role in bridging
    logical connections.
    We believe this does not differ from the edge layer in any significant aspect.

    However, there is a distinction between a \emph{fog layer} as described above and a \emph{fog area}, and that is
    where the term is useful.
    The \emph{fog area} is a geographically--based group of devices, including devices on all layers except the cloud.
    This grouping may be static or dynamic, depending on the properties of the devices---an example is inter--vehicle
    communication, where edge nodes are mobile.
    This grouping allows reasoning about larger--scale device locality which solely edge do not encompass.

    Lastly, there is the cloud layer, which was already described.
    One may note that devices such as routers, are not included, even though they must be present for any meaningful
    infrastructure as described to exist.
    They are a supporting mechanism present on all layers, but are sometimes replaced by alternative connection
    mechanisms that the edge layer provides.

    \section{Levels of management of edge architectures}\label{sec:levels-of-management-of-edge-architectures}
    Apart from the overall architecture, there are also different ways of managing devices and functionalities in edge
    architectures.
    In the cloud, IaaS, PaaS and SaaS are the most common solution types.
    At the edge, no such type has emerged to be the most prominent.

    Similar terms exist in this field. \mbox{Things-aaS\cite{aazam2018fog}} and
    Smart Object-aaS\cite{cavalcante2016interplay} are concepts of exposing sensors, actuators and devices to the
    network by providing managed bridges as an interface between a traditional networked component and \emph{things}
    that may or may not have been originally designed to connect to a network as a managed object.
    Sensing-aaS\cite{perera2013sensing,cavalcante2016interplay} and Sensing and
    Actuation-aaS\cite{merlino2014stack4things,cavalcante2016interplay,aazam2018fog} are similar concepts, but deal
    with exposing the sensing and/or actuation capabilities of devices rather than the devices themselves, providing a
    further abstraction layer between data sources and data consumers.
    Data-aaS, City Infrastructure-aaS are examples of different terms for more or less the same concept, shared between
    all previously mentioned terms: making some kind of data available over the network in a managed system.
    Existing research on this will be presented in Section~\ref{sec:existing-research}.

    The cloud IaaS, PaaS and SaaS concepts can be somewhat extended to the edge.
    With the exception of SaaS, which is focused on end--user applications, concepts of both IaaS and SaaS can be found
    in, for example, Sensing-aaS, where sensors are abstracted and exposed in the same way as computing resources.

    Classically, the main difference between IaaS and PaaS is that the former allows direct use of hardware resources,
    albeit virtualised or somehow isolated, while the latter offers an abstracted development platform and software
    lifecycle and hides the actual underlying hardware\cite{dillon2010cloud}.

    To transfer IaaS and PaaS to the area of edge devices, we need to know the benefits and drawbacks to using devices not
    situated in traditional data centre environments, which offer security, power management and a reliable hardware
    deployment setting.

    Placing devices at the edge, for example in a factory floor or throughout an airport provides no redundancy found
    in a data center---there is only a single power supply with no external management and a single network link with
    potentially low-performing upstream equipment.
    Additionally, physical security is an issue as devices are placed where anyone, potentially even the general
    public, can access them.
    This presents a difficult issue in hardware and network security, as new threat models need to be considered that
    have previously been ignored.

    On the other hand, there are application-specific benefits that placing devices \emph{in situ} bring.
    A factory often lacks the necessary infrastructure for a proper data centre, an airport may have its own data
    centre but require smart sensing devices to analyse data from customers or even manage point--of--sale
    terminals\cite{salis2018benefits}.
    Even with the lack of local device redundancy, in an event of a wider network outage, the local network could be
    retained, offering a limited set of functionalities locally.

    An extension of that is a reduction of decision-making latency that can be achieved by not contacting a distant
    server through a WAN but instead making decisions on the edge device, where network latency can be several orders
    of magnitude lower, enabling applications where real-time decisions are crucial.
    The data security aspect could also be important: privacy constraints could limit data transfer to a cloud service.
    In this case, having devices capable of processing data in a compliant location could be the only way for an
    application to operate.

    \section{Existing research}\label{sec:existing-research}
    Research related to the cloud computing and IoT paradigms is reasonably old, with them being started to be widely
    explored in the mid 2000s.
    The more recent field of fog and edge computing has emerged in the early 2010s.

    On the industrial side, existing providers of cloud platforms have begun to implement and support IoT platforms
    as an extension to their business, attaching additional IoT-- and edge--focused functionalities to their existing
    solutions.
    Due to this, work on edge computing platforms and comparisons between existing solutions has begun to increase.

    Grid computing was considered a predecessor to the IaaS, PaaS and SaaS models.
    Whereas the grid was considered useful for a small number computationally expensive tasks, the benefits of the
    cloud were considered as the capability to provide scalability to a large number of heterogeneous tasks, not
    necessarily compute--intensive.
    A cloud is considered to have the following properties:

    \begin{itemize}
        \item self-service, with no administrator intervention for general usage,
        \item broad, homogenous network access,
        \item resource pooling, multi-tenancy, possibly via virtualisation,
        \item elasticity and scalability and
        \item resource usage measurement, possibly used for billing.
    \end{itemize}

    \subsection{Related work}\label{subsec:related-work}
    Recent work has been mostly focused on integration strategies for platforms.
    Because comparing platforms in--depth is difficult at best, most research does not include any comprehensive
    evaluation, but instead focuses on purely theoretical methods or a simple proof of
    concept\cite{cavalcante2016interplay}.


    Papers appear mostly in workshops or conferences rather than journals, which relates to the relative immaturity
    of the field.
    Publications are spread across around 30 sources~\cite{cavalcante2016interplay}, with no single one seeming
    prevalent.
    More thorough evaluation methods are desired, as 15 \% of evaluations are done on a purely theoretical basis and
    another 40 \% are extremely simple single-purpose proof of concept applications.
    Types of proposals for new platforms can be categorised into the following groups:

    \begin{itemize}
        \item architecture: purely theoretical proposals,
        \item platform: implementations supporting the development and execution of applications in hardware or in software,
        \item framework: software directly used in the development process and
        \item middleware: services applications use.
    \end{itemize}

    The IoT layer is nearly always included\cite{diaz2016stateoftheart} within existing platform as articles---the edge
    is seldom used solely for computation.
    This makes sense, as only relocating computation, without data--generating components near it, brings little
    benefit.

    Approaches to integrating low--level devices are various: one project integrates the sensors directly into existing
    modules in OpenStack\cite{merlino2014stack4things,theopenstackprojectOpenStack}, while another tries to adhere
    to the UNIX philosophy of \emph{everything is a file}, and maps sensors to filesystem
    objects\cite{bruneo2018ocloud}.
    Both resemble IaaS in that they only expose resources as primitives, but do not otherwise provide added services.
    Authors claim language independence and liken their approach to the one in the Raspberry Pi platform, which exposes
    GPIO pins as special filesystem objects.

    In research of the edge and the fog, authors largely equate the two\cite{mahmud2018fog,aazam2018fog}.
    This is done either explicitly, or implicitly without even mentioning the edge layer---using the fog to include its
    functionalities.
    Sources do agree, however, on the features a fog should provide:

    \begin{itemize}
        \item storage, or some kind of persistance mechanism for data,
        \item networking, or a way to connect devices between separate networks and
        \item computational offloading.
    \end{itemize}

    Providing storage can be done in many ways.
    The only important characteristics are that there must be a way to submit and retrieve data to and from the
    solution, as edge devices might not necessarily have the capability to have local instances of databases.

    Networking can be provided either as an overlay network for transparent connectivity--this would provide an
    IaaS--like service.
    There may be a higher--level mechanism for logical connections, akin to PaaS, such as a distributed message queue,
    which applications could explicitly conform to and use.

    An important and often overlooked aspect of edge networking is ensuring reliability.
    Compared to data centres, where one can assume that while reliability of equipment and connections is high,
    at the edge devices are in uncontrolled environments with using reliable hardware.
    Network connections are not fault tolerant and handling this unreliability must be done at the software level,
    or more specifically, at the level of the platform devices are connected to.
    Applications must be able to persist through network connection loss and, ideally, provide functionality even in
    cases with no connection to the global Internet.

    Offloading of computation is necessarily present in fog or edge computation scenarios for them to be useful to
    lower layers.
    How this is implemented is flexible, ranging from grid computing--like solutions, to only spawning whole
    applications on other nodes and subsequently communicating with them, possibly speculatively if there is a demand
    for responsiveness.

    \emph{Data locality} is frequently used in the context of this computation\cite{jiang2018challenges}.
    Used in terms of a single computer, this means cache hits and misses but in a distributed environment that edge
    devices provide, it is used for processing data that resides on the local node, without it being fetched over the
    network.
    Lessons learned from grid distributed computing apply here---not all workloads are necessarily sutable for this.
    Calculating an easily parallelisable task, such as the mean value of a dataset, is simple, however if the
    computation is not trivially parallelisable, challenges due to the relatively high network latency between edge
    nodes arise, which must be solved differently than grid computing problems, which exist in controlled environments.

    The reduction of latency, particularly computation and decision--making, is most often the primary benefit pointed
    out for edge infrastructures~\cite{aazam2018fog}.
    While the balance between data transfer speed and processing speed needs to be achieved, there are use--cases where
    this may be useful.

    \section{Existing solutions}\label{sec:existing-platforms}
    There are a myriad of platforms supporting IoT and edge computing workloads.
    We have selected 32 for an evaluation.
    This will not be an in--depth comparison---the goal is to understand the variety of functionalities, as advertised,
    of this limited set of platforms.
    This is definitely not a comprehensive list of solutions in this niche.
    The products listed have been sourced through browsing review papers, web searches for similar platforms and
    through platforms already known to the authors.

    We have chosen around 25 metrics for comparison.
    These mostly concern categorisation and boolean feature availability and scope, pricing scheme, and general
    popularity.
    This information has been condensed into a table in Figure~\ref{fig:comparison-table}.
    In the following text, we will refer to these platforms by either their name or by their sequential number in the
    first (index) column.

    \begin{figure}
        \centerline{
            \begin{adjustbox}{angle=90}
                \scalebox{0.7}{
                \arrayrulecolor[rgb]{0.8,0.8,0.8}
                \begin{tabular}{rl!{\color{black}\vrule width \lightrulewidth}lllllllllllllllrl}
                    \toprule
                    \multicolumn{1}{l}{} & Provider                                                             & \begin{tabular}[c]{@{}l@{}}Main\\type\end{tabular} & \begin{tabular}[c]{@{}l@{}}Open \\source\end{tabular} & \begin{tabular}[c]{@{}l@{}}Self\\hosted\end{tabular} & \begin{tabular}[c]{@{}l@{}}Provider\\integration\end{tabular} & \begin{tabular}[c]{@{}l@{}}Analytics/\\triggers\end{tabular} & \begin{tabular}[c]{@{}l@{}}Has\\GW\end{tabular} & \begin{tabular}[c]{@{}l@{}}Official \\devices\end{tabular}      & \begin{tabular}[c]{@{}l@{}}Official \\languages\end{tabular}                     & \begin{tabular}[c]{@{}l@{}}Comm \\protocols.\end{tabular}        & \begin{tabular}[c]{@{}l@{}}IoT \\focus\end{tabular} & \begin{tabular}[c]{@{}l@{}}Segment\\focus\end{tabular} & Authn/z                                                       & \begin{tabular}[c]{@{}l@{}}Alt. \\network\end{tabular}       & \begin{tabular}[c]{@{}l@{}}Offline\\funct.\end{tabular} & Pricing                                                                     & \multicolumn{1}{l}{Year} & Docs                                                                 \\
                    \arrayrulecolor{black}\midrule
                    1                    & \begin{tabular}[c]{@{}l@{}}Carriots/Altair \\SmartWorks\end{tabular} & PaaS                                               & no                                                    & optional                                             & no                                                            & yes, yes                                                     & no                                              & no                                                              & Java, Groovy                                                                     & REST, MQTT                                                       & yes                                                 & none                                                   & API key                                                       & \begin{tabular}[c]{@{}l@{}}Sigfox, \\LoRa\end{tabular}       & no                                                      & \begin{tabular}[c]{@{}l@{}}30 day trial, \\quote\end{tabular}               & 2011                     & \begin{tabular}[c]{@{}l@{}}open, \\moderate\end{tabular}             \\
                    \arrayrulecolor[rgb]{0.8,0.8,0.8}\hline
                    2                    & Exosite Murano                                                       & SaaS                                               & no                                                    & optional                                             & no                                                            & yes, yes                                                     & no                                              & Arduino                                                         & Lua                                                                              & \begin{tabular}[c]{@{}l@{}}REST, MQTT,\\ ws\end{tabular}         & yes                                                 & none                                                   & API key                                                       & no                                                           & no                                                      & quote                                                                       & 2009                     & \begin{tabular}[c]{@{}l@{}}open, \\moderate\end{tabular}             \\
                    \hline
                    3                    & Grovestreams                                                         & SaaS                                               & no                                                    & optional                                             & no                                                            & yes, yes                                                     & no                                              & no                                                              & /                                                                                & REST                                                             & yes                                                 & none                                                   & API key                                                       & no                                                           & no                                                      & \begin{tabular}[c]{@{}l@{}}free tier,\\per I/O, users\end{tabular}          & 2011                     & \begin{tabular}[c]{@{}l@{}}open, \\very limited\end{tabular}         \\
                    \hline
                    4                    & \begin{tabular}[c]{@{}l@{}}Realtime.io/\\ioBridge\end{tabular}       & SaaS                                               & no                                                    & no                                                   & no                                                            & yes, yes                                                     & no                                              & \begin{tabular}[c]{@{}l@{}}proprietary, \\Arduino\end{tabular}  & JS                                                                               & REST                                                             & yes                                                 & none                                                   & API key                                                       & no                                                           & no                                                      & quote                                                                       & 2008                     & \begin{tabular}[c]{@{}l@{}}open, \\very limited\end{tabular}         \\
                    \hline
                    5                    & Sensorcloud                                                          & SaaS                                               & no                                                    & no                                                   & no                                                            & yes, yes                                                     & no                                              & proprietary                                                     & Java, C\#                                                                        & REST                                                             & yes                                                 & none                                                   & API key                                                       & yes                                                          & no                                                      & \begin{tabular}[c]{@{}l@{}}free tier, per \\transaction, alert\end{tabular} & 2011                     & \begin{tabular}[c]{@{}l@{}}open, \\moderate\end{tabular}             \\
                    \hline
                    6                    & Tempoiq                                                              & SaaS                                               & no                                                    & no                                                   & no                                                            & yes, yes                                                     & no                                              & no                                                              & \begin{tabular}[c]{@{}l@{}}Python, JS, \\Ruby, Java, C\#\end{tabular}            & REST, MQTT                                                       & yes                                                 & none                                                   & API key                                                       & no                                                           & no                                                      & trial, quote                                                                & 2016                     & \begin{tabular}[c]{@{}l@{}}open, \\limited\end{tabular}              \\
                    \hline
                    7                    & Thingworx                                                            & SaaS                                               & no                                                    & optional                                             & no                                                            & yes, yes                                                     & no                                              & \begin{tabular}[c]{@{}l@{}}Android, \\iOS\end{tabular}          & \begin{tabular}[c]{@{}l@{}}C, C\#, \\Java, ObjC\end{tabular}                     & REST, MQTT                                                       & no                                                  & \begin{tabular}[c]{@{}l@{}}Industry\\4.0\end{tabular}  & \begin{tabular}[c]{@{}l@{}}API key, \\LDAP\end{tabular}       & no                                                           & no                                                      & trial, quote                                                                & 2014                     & videos                                                               \\
                    \hline
                    8                    & Wotkit                                                               & SaaS                                               & no                                                    & no                                                   & no                                                            & yes, no                                                      & no                                              & no                                                              & /                                                                                & REST                                                             & yes                                                 & none                                                   & \begin{tabular}[c]{@{}l@{}}user/pass,\\ OAuth2\end{tabular}   & no                                                           & no                                                      & quote                                                                       & 2010                     & \begin{tabular}[c]{@{}l@{}}open, \\very limited\end{tabular}         \\
                    \hline
                    9                    & Lelylan                                                              & PaaS                                               & yes                                                   & yes                                                  & no                                                            & no, yes                                                      & no                                              & no                                                              & /                                                                                & \begin{tabular}[c]{@{}l@{}}REST, MQTT,\\ ws\end{tabular}         & yes                                                 & none                                                   & API key                                                       & no                                                           & no                                                      & open source                                                                 & 2011                     & \begin{tabular}[c]{@{}l@{}}open, \\moderate\end{tabular}             \\
                    \hline
                    10                   & \begin{tabular}[c]{@{}l@{}}Mathworks \\Thingspeak\end{tabular}       & SaaS                                               & yes                                                   & optional                                             & yes                                                           & yes, yes                                                     & no                                              & no                                                              & C, Python                                                                        & REST, MQTT                                                       & yes                                                 & none                                                   & API key                                                       & no                                                           & no                                                      & trial, per I/O                                                              & 2010                     & \begin{tabular}[c]{@{}l@{}}open, \\extensive\end{tabular}            \\
                    \hline
                    11                   & Stack4Things                                                         & IaaS                                               & yes                                                   & yes                                                  & no                                                            & no, no                                                       & yes                                             & no                                                              & /                                                                                & REST, CoAP                                                       & no                                                  & none                                                   & API key                                                       & no                                                           & no                                                      & open source                                                                 & 2014                     & \begin{tabular}[c]{@{}l@{}}open, \\very limited\end{tabular}         \\
                    \hline
                    12                   & C3 IoT                                                               & SaaS                                               & no                                                    & no                                                   & yes                                                           & yes, n/a                                                     & no                                              & no                                                              & n/a                                                                              & n/a                                                              & yes                                                 & \begin{tabular}[c]{@{}l@{}}Industry \\4.0\end{tabular} & n/a                                                           & no                                                           & no                                                      & quote                                                                       & 2016                     & closed                                                               \\
                    \hline
                    13                   & Parse Platform                                                       & PaaS                                               & yes                                                   & yes                                                  & no                                                            & yes, yes                                                     & no                                              & no                                                              & \begin{tabular}[c]{@{}l@{}}ObjC, JS, Java, \\.NET, PHP\end{tabular}              & REST                                                             & yes                                                 & none                                                   & API key                                                       & no                                                           & no                                                      & open source                                                                 & 2011                     & \begin{tabular}[c]{@{}l@{}}open, \\extensive\end{tabular}            \\
                    \hline
                    14                   & \begin{tabular}[c]{@{}l@{}}Salesforce \\IoT\end{tabular}             & SaaS                                               & no                                                    & no                                                   & yes                                                           & yes, n/a                                                     & no                                              & no                                                              & /                                                                                & REST                                                             & yes                                                 & Salesforce                                             & OAuth2                                                        & no                                                           & no                                                      & quote                                                                       & 2015                     & closed                                                               \\
                    \hline
                    15                   & \begin{tabular}[c]{@{}l@{}}Oracle \\IoT Cloud\end{tabular}           & PaaS                                               & no                                                    & no                                                   & yes                                                           & yes, yes                                                     & yes                                             & no                                                              & \begin{tabular}[c]{@{}l@{}}JS, Java, \\ObjC, C\end{tabular}                      & REST                                                             & yes                                                 & none                                                   & OAuth2                                                        & no                                                           & no                                                      & per cpu/hour                                                                & 2017                     & \begin{tabular}[c]{@{}l@{}}open, \\extensive\end{tabular}            \\
                    \hline
                    16                   & Kaa                                                                  & SaaS                                               & yes                                                   & optional                                             & no                                                            & yes, yes                                                     & yes                                             & no                                                              & \begin{tabular}[c]{@{}l@{}}C++, ObjC, \\Java\end{tabular}                        & MQTT, CoAP                                                       & yes                                                 & none                                                   & user/pass                                                     & no                                                           & no                                                      & \begin{tabular}[c]{@{}l@{}}trial, hosted,\\ self-hosted\end{tabular}        & 2014                     & \begin{tabular}[c]{@{}l@{}}open, \\extensive\end{tabular}            \\
                    \hline
                    17                   & Temboo                                                               & PaaS                                               & no                                                    & no                                                   & no                                                            & yes, yes                                                     & no                                              & Arduino                                                         & \begin{tabular}[c]{@{}l@{}}Java, PHP, C\#, \\ObjC, py, JS, Ruby\end{tabular}     & \begin{tabular}[c]{@{}l@{}}REST, MQTT,\\ CoAP\end{tabular}       & yes                                                 & \begin{tabular}[c]{@{}l@{}}Industry \\4.0\end{tabular} & \begin{tabular}[c]{@{}l@{}}user/pass, \\OAuth2\end{tabular}   & no                                                           & no                                                      & trial, quote                                                                & 2006                     & \begin{tabular}[c]{@{}l@{}}open, \\extensive\end{tabular}            \\
                    \hline
                    18                   & \begin{tabular}[c]{@{}l@{}}Ayla \\IoT Fabric\end{tabular}            & PaaS                                               & no                                                    & no                                                   & no                                                            & yes, yes                                                     & yes                                             & RTOS                                                            & n/a                                                                              & n/a                                                              & no                                                  & none                                                   & n/a                                                           & \begin{tabular}[c]{@{}l@{}}BLE, Z-Wave,\\Zigbee\end{tabular} & no                                                      & quote                                                                       & 2010                     & closed                                                               \\
                    \hline
                    19                   & ThethingsIO                                                          & PaaS                                               & partial                                               & optional                                             & no                                                            & yes, yes                                                     & no                                              & no                                                              & /                                                                                & REST, MQTT                                                       & yes                                                 & none                                                   & API key                                                       & \begin{tabular}[c]{@{}l@{}}LoRa, NB-IoT\\Sigfox\end{tabular} & no                                                      & \begin{tabular}[c]{@{}l@{}}trial, quote, per\\ device/message\end{tabular}  & 2013                     & \begin{tabular}[c]{@{}l@{}}open, \\extensive\end{tabular}            \\
                    \hline
                    20                   & OpenRemote                                                           & PaaS                                               & yes                                                   & yes                                                  & no                                                            & no, yes                                                      & yes                                             & no                                                              & /                                                                                & \begin{tabular}[c]{@{}l@{}}REST, SNMP, \\1wire\end{tabular}      & yes                                                 & \begin{tabular}[c]{@{}l@{}}smart \\home\end{tabular}   & API key                                                       & no                                                           & yes                                                     & \begin{tabular}[c]{@{}l@{}}open source, \\quote\end{tabular}                & 2016                     & \begin{tabular}[c]{@{}l@{}}open, \\extensive\end{tabular}            \\
                    \hline
                    21                   & \begin{tabular}[c]{@{}l@{}}Cisco \\Jasper\end{tabular}               & IaaS                                               & no                                                    & no                                                   & no                                                            & yes, yes                                                     & no                                              & n/a                                                             & n/a                                                                              & n/a                                                              & no                                                  & none                                                   & n/a                                                           & NB-IoT                                                       & no                                                      & quote                                                                       & 2004                     & closed                                                               \\
                    \hline
                    22                   & Cumulocity                                                           & PaaS                                               & no                                                    & gateway                                              & no                                                            & yes, yes                                                     & yes                                             & a lot                                                           & \begin{tabular}[c]{@{}l@{}}Java, C++,\\C\#\end{tabular}                          & REST, MQTT                                                       & no                                                  & none                                                   & user/pass                                                     & \begin{tabular}[c]{@{}l@{}}LoRa, \\Sigfox\end{tabular}       & yes                                                     & \begin{tabular}[c]{@{}l@{}}trial, quote, \\per device\end{tabular}          & 2010                     & \begin{tabular}[c]{@{}l@{}}open, \\extensive\end{tabular}            \\
                    \hline
                    23                   & Cloudplugs                                                           & PaaS                                               & partial                                               & optional                                             & no                                                            & yes, yes                                                     & no                                              & \begin{tabular}[c]{@{}l@{}}Arduino, \\ESP32\end{tabular}        & \begin{tabular}[c]{@{}l@{}}Java, C, JS, \\ObjC, Python\end{tabular}              & \begin{tabular}[c]{@{}l@{}}REST, MQTT, \\ws\end{tabular}         & yes                                                 & \begin{tabular}[c]{@{}l@{}}Industry \\4.0\end{tabular} & user/pass                                                     & \begin{tabular}[c]{@{}l@{}}LoRa, BLE,\\Z-Wave\end{tabular}   & no                                                      & quote                                                                       & 2014                     & \begin{tabular}[c]{@{}l@{}}partially open, \\extensive\end{tabular}  \\
                    \hline
                    24                   & FIWARE                                                               & PaaS                                               & yes                                                   & yes                                                  & no                                                            & no, yes                                                      & no                                              & no                                                              & /                                                                                & REST, MQTT                                                       & yes                                                 & none                                                   & API key                                                       & no                                                           & no                                                      & open source                                                                 & 2011                     & \begin{tabular}[c]{@{}l@{}}open, \\limited\end{tabular}              \\
                    \hline
                    25                   & OpenMTC                                                              & PaaS                                               & yes                                                   & yes                                                  & no                                                            & yes, yes                                                     & no                                              & no                                                              & /                                                                                & REST, MQTT                                                       & no                                                  & none                                                   & API key                                                       & Zigbee                                                       & no                                                      & open source                                                                 & 2017                     & \begin{tabular}[c]{@{}l@{}}open, \\very limited\end{tabular}         \\
                    \hline
                    26                   & Sitewhere                                                            & SaaS                                               & yes                                                   & yes                                                  & no                                                            & yes, yes                                                     & no                                              & no                                                              & /                                                                                & \begin{tabular}[c]{@{}l@{}}REST, MQTT, \\AMQP, CoAP\end{tabular} & yes                                                 & none                                                   & API key                                                       & no                                                           & no                                                      & \begin{tabular}[c]{@{}l@{}}community, \\quote\end{tabular}                  & 2010                     & \begin{tabular}[c]{@{}l@{}}open, \\limited\end{tabular}              \\
                    \hline
                    27                   & Kura                                                                 & PaaS                                               & yes                                                   & yes                                                  & no                                                            & no, no                                                       & no                                              & no                                                              & Java                                                                             & REST, MQTT                                                       & yes                                                 & none                                                   & API key                                                       & no                                                           & no                                                      & open source                                                                 & 2013                     & \begin{tabular}[c]{@{}l@{}}open, \\moderate\end{tabular}             \\
                    \hline
                    28                   & Node-RED                                                             & PaaS                                               & yes                                                   & yes                                                  & no                                                            & yes, yes                                                     & no                                              & \begin{tabular}[c]{@{}l@{}}Arduino, \\rPi, Android\end{tabular} & visual, JS                                                                       & REST, MQTT                                                       & yes                                                 & \begin{tabular}[c]{@{}l@{}}smart \\home\end{tabular}   & API key                                                       & no                                                           & no                                                      & open source                                                                 & 2013                     & \begin{tabular}[c]{@{}l@{}}open, \\extensive\end{tabular}            \\
                    \hline
                    29                   & \begin{tabular}[c]{@{}l@{}}IBM \\Watson IoT\end{tabular}             & SaaS                                               & no                                                    & no                                                   & yes                                                           & yes, yes                                                     & no                                              & n/a                                                             & n/a                                                                              & n/a                                                              & yes                                                 & none                                                   & n/a                                                           & no                                                           & no                                                      & \begin{tabular}[c]{@{}l@{}}per instance, \\feature, quote\end{tabular}      & 2015                     & closed                                                               \\
                    \hline
                    30                   & AWS IoT                                                              & PaaS                                               & no                                                    & gateway                                              & yes                                                           & yes, yes                                                     & yes                                             & a lot                                                           & \begin{tabular}[c]{@{}l@{}}Java, .NET, JS, PHP,\\ py, Ruby, Go, C++\end{tabular} & \begin{tabular}[c]{@{}l@{}}REST, MQTT, \\ws\end{tabular}         & no                                                  & none                                                   & cert                                                          & no                                                           & yes                                                     & \begin{tabular}[c]{@{}l@{}}per call, device, \\action, tx\end{tabular}      & 2017                     & \begin{tabular}[c]{@{}l@{}}open, \\extensive\end{tabular}            \\
                    \hline
                    31                   & \begin{tabular}[c]{@{}l@{}}GCP IoT\\/Xively\end{tabular}             & PaaS                                               & no                                                    & gateway                                              & yes                                                           & yes, yes                                                     & yes                                             & a lot                                                           & \begin{tabular}[c]{@{}l@{}}Java, JS, \\py, C++\end{tabular}                      & REST, MQTT                                                       & no                                                  & none                                                   & cert                                                          & no                                                           & no                                                      & \begin{tabular}[c]{@{}l@{}}per call, \\data, tx\end{tabular}                & 2017                     & \begin{tabular}[c]{@{}l@{}}open, \\moderate\end{tabular}             \\
                    \hline
                    32                   & Azure IoT                                                            & PaaS                                               & partial                                               & gateway                                              & yes                                                           & yes, yes                                                     & yes                                             & a lot                                                           & \begin{tabular}[c]{@{}l@{}}Java, .NET,\\JS, py, C\end{tabular}                   & \begin{tabular}[c]{@{}l@{}}REST, MQTT, \\AMQP\end{tabular}       & no                                                  & none                                                   & \begin{tabular}[c]{@{}l@{}}API key, \\cert, SASL\end{tabular} & no                                                           & yes                                                     & \begin{tabular}[c]{@{}l@{}}per unit, action, \\message, device\end{tabular} & 2016                     & \begin{tabular}[c]{@{}l@{}}open, \\very extensive\end{tabular}       \\
                    \bottomrule
                \end{tabular}
                \arrayrulecolor{black}
                }
            \end{adjustbox}
        }
        \caption{Summary platform comparison table.}
        \label{fig:comparison-table}
    \end{figure}

    A basic comparison is the type of the platform, classified into IaaS, PaaS and SaaS\@.
    Existing cloud provider features can easily be categorised into these three groups, but this is not the case for
    edge computing platforms, where there is not much variety.
    Essentially, most platforms use a PaaS model, providing tools to the developer to explicitly use when developing,
    often resulting in vendor lock--in.
    SaaS platforms exist for end--user solutions and IaaS platforms are the rarest.
    The platform focused most on the style of IaaS was Cisco Jasper\cite{ciscojasper}, which focuses on device
    connectivity.

    Most platforms are deployed as a service managed by the provider, with some being available to completely
    self--host.
    The vast majority of open--source components are able to be self--hosted, and about half of the others offer the
    ability to host a component on a private infrastructure, connecting to the global cloud deployment.

    Even the platforms that offer edge computing devices (11, 15, 16, 18, 20, 22, 30, 31, 32) or software that connects
    to an existing cloud only allow for a single additional layer of devices.
    Using the layers defined in~\ref{sec:commonly-used-terms}, the cloud layer is always available, with the
    optional gateways acting as the edge layer as an intermediary to IoT devices.
    There is no platform that offers a variable number of layers, or the ability to have more than three layers.

    Integrations for platforms from providers with existing cloud solutions (10, 12, 14, 15, 29, 30, 31, 32) are mostly
    for inventory and access management and data pipelines.
    This enables processing data and integrating into existing applications using the wider cloud platform, but almost
    all solutions offer analytics, triggers and a web dashboard out--of--the--box.
    An interesting exception is MathWorks ThingSpeak\cite{thingspeak}, which offers data processing through MATLAB,
    an existing desktop product instead.

    Most platforms, except (7, 11, 18, 21, 22, 25, 30, 31, 32), focus solely on IoT, which means only focusing on
    acquiring and processing sensor data, either without or with limited ability to run other computation or
    applications on the platform.
    About a third are generalised to be able to operate under the edge computing paradigm to varying degrees---these
    are mostly the ones also offering an edge gateway solution.
    Within the IoT-focused frameworks, there is not often a focus on a specific segment of the industry, but about a
    quarter do: mostly focusing on targeting Industry 4.0, with some also explicitly targeting the smart home market.

    \subsection{Technical details}\label{subsec:technical-details}
    Around half of the platforms (see the \emph{Official devices} column in Figure~\ref{fig:comparison-table}) offer
    some kind of explicit support for IoT devices in the form of usage tutorials or real--time operating system
    support.
    The most commonly supported platforms are Arduino and Raspberry Pi, with larger or more focused industrial
    providers also supporting more specialised devices.
    Frameworks excelling in this category are Cumulocity\cite{cumulocity}, AWS IoT\cite{awsiot},
    Google Cloud Platform IoT\cite{gcpiot} and Azure IoT\cite{azureiot}, particularly the latter, offering an
    exceptionally large number of devices with software and hardware integrations.

    All platforms offer a programming language-agnostic way to interface with the platform with an HTTP API\@.
    Other \emph{official} programming languages differ by platform, but Java, Python, C, C\# and Javascript are most
    frequently supported.
    The presence of SDKs and examples are correlated with platform popularity.

    The most common mechanism for securing, authenticating and authorising transmissions is the combination of an API
    key along with TLS encryption.
    This is sometimes used in an OAuth2 context or, less frequently, with HTTP Basic authentication.
    A common configuration is per--device X.509 certificates, serving the dual purpose of inventory management.

    Alternative networking protocols are seldom supported.
    They are, in order of decreasing frequency of support: LoRa, Sigfox, ZigBee, Bluetooth, Z-Wave and NB-IoT\@.
    Gateway devices offer this connection bridging capability, particularly for transmitting sensor data.
    Kaa IoT\cite{kaaiot} is the only platform offering explicit support for battery management by batching updates,
    while AWS IoT and Azure IoT stand out with explicit support for intermittent connectivity, offering a subset of
    functionalities locally.

    \subsection{Pricing and popularity}\label{subsec:pricing-and-popularity}
    Pricing varies greatly between the platforms.
    Open source components are offered for free or with paid plans for hosted solutions, while others have various
    methods of managing costs.
    Some have bulk packages with quotas, others have prices scaling with the number of connected devices.
    Typical of platforms provided by companies with existing cloud services are very verbose pricing plans, charging by
    the number of actions performed, API calls or bytes transferred in very small increments.
    A number of the projects, particularly those not generally popular, do not have public pricing plans, instead
    requiring a direct contact for a pricing inquiry.

    We measured popularity through Google Trends and categorised platforms into 5 groups of popularity based on
    their current or historical popularity and growth rate.
    The groups are descriptive with their members being fairly similar among themselves, but their popularity
    quantifier is subjective.
    We make no claim to the quality of the platforms through this metric, but recognise that community support and the
    availability of documentation are very important for development.
    The categories are, excluding platforms which do not appear on Google Trends at all:

    \begin{enumerate}
        \item very high popularity: Node-RED, AWS IoT, GCP IoT, Azure IoT
        \item high popularity: Thingworx, Mathworks Thingspeak, FIWARE
        \item low popularity: ioBridge, C3 IoT, Salesforce IoT, Temboo, OpenRemote, Cisco Jasper, Cumulocity
        \item very low popularity: Oracle IoT Cloud, Kaa, Sitewhere
        \item extremely low popularity: Grovestreams, Tempoiq, Lelylan, thethings.io, Cloudplugs
    \end{enumerate}


    There are a few platforms that stand out from others in particular aspects.
    As mentioned before, Cisco Jasper is the only platform focused purely on infrastructure management.
    Ayla IoT Fabric\cite{aylaiot} and OpenRemote\cite{openremote} are the only frameworks offering data sharing between
    users---the idea being that multiple users connect their devices into a wider network, giving each themselves the
    possibility of selectively sharing sensor data with other users.

    AWS IoT, GCP IoT and Azure IoT are seemingly the most mature and popular products, objectively offering the most
    features, with integrations into the wider platforms of their respective providers.

    \section{Conclusion}\label{sec:conclusion}
    We have discussed to different approaches to understanding cloud, fog, edge and IoT architectures, reviewed
    relevant literature and investigated the platforms currently available on the market.

    The differences in the interpretation and understanding of especially the terms of \emph{fog} and \emph{edge} were
    large, as different sources place functionalities into different groups.
    These overlap, so there is no definitive agreement on a precise definition of the terms, but we have managed to
    identify key features sources use when referencing them.

    Levels of conformance to established IaaS/PaaS/SaaS styles were also considered, finding a large overlap but not a
    definitive mapping.
    In existing literature, there is a large variety of approaches to building new systems.
    Evaluating them leaves much to be desired, though, as comparisons are frequently very shallow.

    Existing solutions do not cover the area of fog computing, but some support for edge computing is present,
    frequently in the form of edge gateways supporting delegating functionality from the cloud.
    We have identified the key characteristics of a multitude of commercial and open--source platforms and found that
    there are clear leaders in functionalities, but there exist leaders in specific niches targeting specific needs.

    Our work is ongoing with these being our initial results.
    In the future, we plan to make a more detailed and methodological comparison, with a PoC implementation in some
    of the most platforms.

    \pagebreak

    \bibliographystyle{splncs04}
    \bibliography{literature}
\end{document}